\documentclass[pre,twocolumn,aps,superscriptaddress,showpacs,floatfix]{revtex4}

\usepackage{amsmath}
\usepackage{amssymb}
\usepackage{amsbsy}
\usepackage{graphicx}
\usepackage{color}
 
\usepackage{enumerate}
\usepackage{mathtools}

\newcommand{\ee}[0]{\boldsymbol{e}}

\newcommand{\ii}[0]{\mathrm{i}}

\newcommand{\At}[0]{\widetilde{A}}

\newcommand{\ex}[1]{\mathrm{e}^{#1}}

\newcommand{\emuu}[0]{\mathbf{e_{\boldsymbol{\mu}}}}
\newcommand{\lamb}[0]{\boldsymbol{\lambda}}
\newcommand{\enu}[0]{\mathbf{e_{\boldsymbol \nu}}}

\newcommand{\dd}[0]{\mathrm{d}}

\begin{document}

\title{Nonlinear response and emerging nonequilibrium micro-structures for biased diffusion in confined crowding environments}

\author{O. B\'enichou}
\affiliation{Laboratoire de Physique Th\'eorique de la Mati\`ere Condens\'ee, UPMC, CNRS UMR 7600,
Sorbonne Universit\'es, 4 Place Jussieu, 75252 Paris Cedex 05, France}

\author{P. Illien}
\affiliation{Laboratoire de Physique Th\'eorique de la Mati\`ere Condens\'ee, UPMC, CNRS UMR 7600,
Sorbonne Universit\'es, 4 Place Jussieu, 75252 Paris Cedex 05, France}
\affiliation{Rudolf Peierls Centre for Theoretical Physics, University of Oxford, Oxford OX1 3NP, UK}
\affiliation{Department of Chemistry, The Pennsylvania State University, University Park, Pennsylvania 16802, USA}

\author{G. Oshanin}
\affiliation{Laboratoire de Physique Th\'eorique de la Mati\`ere Condens\'ee, UPMC, CNRS UMR 7600,
Sorbonne Universit\'es, 4 Place Jussieu, 75252 Paris Cedex 05, France}

\author{A. Sarracino}
\affiliation{Laboratoire de Physique Th\'eorique de la Mati\`ere Condens\'ee, UPMC, CNRS UMR 7600,
Sorbonne Universit\'es, 4 Place Jussieu, 75252 Paris Cedex 05, France}

\author{R. Voituriez}
\affiliation{Laboratoire de Physique Th\'eorique de la Mati\`ere Condens\'ee, UPMC, CNRS UMR 7600,
Sorbonne Universit\'es, 4 Place Jussieu, 75252 Paris Cedex 05, France}

\begin{abstract}
  We study analytically the dynamics and the micro-structural changes
  of a host medium caused by a driven tracer particle moving in a
  confined, quiescent molecular crowding environment. Imitating
  typical settings of active micro-rheology experiments, we consider
  here a minimal model comprising a geometrically confined lattice
  system -- a two-dimensional strip-like or a three-dimensional
  capillary-like -- populated by two types of hard-core particles with
  stochastic dynamics -- a tracer particle driven by a constant
  external force and bath particles moving completely at random.
  Resorting to a decoupling scheme, which permits us to go beyond the
  linear-response approximation (Stokes regime) for arbitrary
  densities of the lattice gas particles, we determine the
  force-velocity relation for the tracer particle and the stationary
  density profiles of the host medium particles around it.  These
  results are validated \emph{a posteriori} by extensive numerical
  simulations for a wide range of parameters. Our theoretical analysis
  reveals two striking features: a) We show that, under certain
  conditions, the terminal velocity of the driven tracer particle is a
  nonmonotonic function of the force, so that in some parameter range
  the differential mobility becomes negative, and b) the biased
  particle drives the whole system into a nonequilibrium steady-state
  with a stationary particle density profile past the tracer, which
  decays exponentially, in sharp contrast with the behavior observed
  for unbounded lattices, where an algebraic decay is known to take
  place.
\end{abstract}


\maketitle

\section{Introduction}

Rheological properties of soft matter systems at the micro-scale can
be investigated by studying the motion of an active tracer particle
(TP), or some other probe, that can sample length scales typical of
the microstructure of the host medium. The study of the system
response to a local perturbation is particularly relevant to complex
and heterogeneous media, such as glasses, gels, and many biological
systems.  Experimentally, the microscopic probe can be driven
externally by using magnetic or optical tweezers; such an experimental
technique, called active micro-rheology, (where the term ``active''
emphasizes that the tracer is not in equilibrium with the environment
but rather induces some micro-structural changes in the host medium),
has been successfully applied to probe the micro-rheological
properties of a variety of different systems (see, e.g.,
\cite{SM10,WP11} for recent reviews), including living
cells~\cite{WMT06}, colloidal suspensions~\cite{PV14}, soft glassy
materials~\cite{GCOAB08,JMCBC12}, and granular
media~\cite{CD10,WGZS14,GPSV14,SGPV15}, to name but a few.

It was realized, in particular, that in all these systems the presence
of a boundary and interactions with it, as well as the specific
geometry of the sample affect not only the dynamics of the tracer
particle but are also relevant for the response properties of the
medium. Indeed, the study of driven diffusion in confined
geometries~\cite{BHMST08}, e.g., in micro-channels, where the
surface-to-volume ratios are large, opened up a broad front of
research, with important applications in several fields, such as,
e.g., micro- and nanofluidics~\cite{MG01,SHR08}, hydrodynamics at
solid-liquid interfaces~\cite{JYB06}, reaction-diffusion
kinetics~\cite{BCKMV10}, dynamics of
colloids~\cite{B02,GKKRWH08,TM08,AR11} and polymers~\cite{G11}, dense
liquids~\cite{LF14}, and crowded systems~\cite{WHVB12,BBCILMOV13}.

In order to characterize the dynamics of the TP, biased by a constant
force $F$, and its effect on the host medium, the first goal is to
establish the so-called force-velocity relation $V(F)$; namely, the
dependence of the stationary velocity $V$ of the probe on the applied
force. This characteristic curve depends on the mutual interaction
between the tracer and the surrounding medium.  In the context of
active micro-rheology of colloidal suspensions, different theoretical
approaches to relate the tracer dynamics with the structural
properties of the host medium have been proposed: in the low density
limit, the friction coefficient in the nonlinear response regime can
be obtained from a simplified many-body Smoluchowski equation, where
the knowledge of the density profile around the tracer is
required~\cite{SB05}; in the dense limit, a mode-coupling
approximation allows one to compute the friction coefficient from the
probe's position correlation function~\cite{FC02}.  A general finding
is that $V(F)$ can show non-trivial nonlinear behaviors, such as
force-thinning or thickening~\cite{PV14,WS15}.

A second important point in the characterization of the system
response is to understand how the motion of the TP perturbs the
surrounding medium and whether this perturbation is localized or, on
the contrary, is long-ranged.  This information is contained in the
particle density profiles around the tracer.  Due to the driving force
acting on the TP, one expects to observe an accumulation of particles
in front of it -- a ``traffic jam'', and a low-density wake, depleted
by the bath particles behind it. This traffic jam produces an extra
contribution to the frictional force exerted on the TP, resulting in
an increased friction~\cite{BCCMO00}.  Clearly, the unperturbed value of the
equilibrium particle density should be approached at infinitely large
distances from it. Such asymmetric density profiles are observed, for
instance, in simulations of colloidal dispersions~\cite{CB05,ZB10} or
lattice gases~\cite{MO10}, in experiments~\cite{SMF10}, and in
glass-forming soft-sphere mixtures~\cite{WH13,DBJ14}.  This
inhomogeneous spatial distribution of the bath particles induced by
the TP and traveling along together with it signifies that the TP
drives the whole system out of equilibrium bringing it to a
nonequilibrium steady-state. This has important physical
consequences, e.g., it can give rise to effective bath- and
bias-mediated interactions in the system~\cite{DLL03,MO10}.
Quantitatively, an exact calculation of the density profiles around
stationary moving tracer particle is impossible, since here one faces
a genuine essentially many-body problem and one has to resort to
approximate approaches.  In particular, for colloidal suspensions
important results have been obtained using the dynamic density
functional theory~\cite{PDT03,RDKP07,TM08}.

Analytical results describing the nonlinear coupling between the TP
and the dynamical environment can be obtained within the framework of
lattice gases -- minimalistic models that are able to reproduce some
of the main features of more complex systems. 
Lattice gas models consist of particles that can jump from one site of
the lattice to another, with the only constraint that each site can be
occupied at most by one particle (excluded-volume interactions). This
physically means that apart from constraining particle dynamics on the
lattice, we take into account only the repulsive part of the
particle-particle interaction potential.  The host medium is
represented by bath particles, performing a (symmetric in all
directions) random walk among the neighboring lattice sites.  The TP
is also considered as a mobile hard-core particle, equal in size to
the bath particles, but may have a different characteristic jump time
and, in active micro-rheology settings, be subjected to a constant
external force favoring its jumps in a preferential direction, see
Fig.~\ref{fig0}, in agreement with the local detailed balance
(LDB)~\cite{LS99}. The model can then be viewed as an asymmetric
exclusion process (the so-called ASEP) evolving in a sea of symmetric
exclusion processes (SEPs)~\cite{M15}, and represents a combination of
two paradigmatic models for transport phenomena in nonequilibrium
statistical mechanics.

Recently, the behavior of $V(F)$ beyond the linear regime in lattice
gas models has received great attention, triggered by the observation
of the striking phenomenon of negative differential mobility (NDM),
namely a nonmonotonic dependence of the TP velocity on the applied
force~\cite{P05,JKGC08,LF13,BBMS13,BM14,BIOSV14,BSV15}. Upon a gradual
increase of the force $F$, the velocity grows linearly, as prescribed
by the linear response, then approaches a maximal value and further
decreases with an increase of $F$. This remarkable ``getting more from
pushing less'' behavior~\cite{ZPM02} occurs not only in lattice
models, but it is observed in several systems, such as nonequilibrium
steady states~\cite{ZPM02}, Brownian motors~\cite{CM96,KMHLT06}, and
kinesin models~\cite{AWV15}. In particular, in the context of
kinetically constraint models, NDM has been observed in numerical
simulations in~\cite{JKGC08,S08} and related to the heterogeneity and
intermittency of the dynamics in the glassy phase~\cite{JKGC08}. More
recently, an analytical theory accounting for NDM in a Lorentz lattice
gas, where the TP travels among fixed obstacles, has been presented by
Leitmann and Franosch~\cite{LF13} in the dilute limit (i.e. at linear
order in the obstacle density). Later, Basu and Maes~\cite{BM14}
observed via numerical simulations the same phenomenon in a dynamical
environment, where obstacles can diffuse with excluded-volume
interactions, and related NDM with the ``frenetic'' contributions
appearing in a nonequilibrium fluctuation-dissipation
relation~\cite{LCZ05,BMW09,LCSZ08,BKLM15,BM15}. A general analytical
approach, accounting for NDM in dynamical environments, and valid for
arbitrary density and arbitrary choice of transition rates, has been
then proposed and discussed in~\cite{BIOSV14}, where the trapping
effect induced on the TP by the coupling between the density and the
characteristic time scale of bath particles, was indicated as the main
physical mechanism responsible for the phenomenon.  The study of the
dependence of NDM on the microscopic dynamical rules has been further
investigated by Baiesi and coworkers~\cite{BSV15}, who stressed that
the particular shape of the obstacles and their coupling with the TP
can play a central role in determining the effective trapping
mechanism.

In this paper we present a general analytical treatment which unifies
and extends previous studies, and allows us to compute the
force-velocity relation of the TP and the density profiles around it
for arbitrary densities and for arbitrary values of the exerted force,
in experimentally relevant \emph{geometrically-restricted systems} --
strip-like or capillary-like confinement; namely, for bounded lattices
which have an infinite extent in the direction of the applied force
and are finite in other directions, perpendicular to the direction of
the applied force, see Fig.~\ref{fig1}. Our theoretical approach is
based on a decoupling of correlation functions of the site occupation
variable and can be applied to any choice of the transition rates,
including, in particular, the cases studied in the recent literature
mentioned above.  This approximation has been previously introduced
for one-dimensional models in~\cite{BOMM92,BOMR96,BCLMO99} and
subsequently generalized for higher dimensional infinite lattices
in~\cite{BCMO99,BCCMO00,BCCMO01}.  Here we extend this analysis over
an experimentally-relevant, but technically more involved case of
dense crowded environments placed in confined,
geometrically-restricted systems. As mentioned above, these particular
boundary conditions have important applications in many systems and
therefore it is worth to study how they influence the system's
response.

Our first main result is the derivation of the analytic expression for
the \emph{velocity} $V(F)$ from the decoupling approximation, valid
for any choice of jump probabilities and arbitrary density.  The
comparison of this result with extensive numerical simulations reveals
that the approximation is very accurate in a wide range of parameters.
We proceed to show that, in a certain region of the parameter space
but not always, the force-velocity relation is a nonmonotonous
function, so that when $F$ exceeds a certain value, at which $V$
attains a maximum, the velocity starts to decrease. In this regime,
remarkably, the differential mobility becomes negative.  Moreover, we
study in detail the behaviors arising from two specific choices of the
TP jump probabilities, considered recently
in~\cite{LF13,BM14,BIOSV14}, both satisfying the LDB, but differing in
the explicit dependence on $F$ in the directions orthogonal to the
force. We derive the phase chart in the parameter space of the model
where NDM is expected to occur for both kinds of jump probabilities.
Our study unifies and clarifies quantitatively the recent analyses
mentioned above, on the role of the microscopic dynamical rules in
nonequilibrium systems, driven beyond the linear response
regime~\cite{BIOSV14,BSV15}.

The second issue we address concerns the effect of boundary conditions
on the \emph{density profiles} around the TP.  In particular, it is
unclear in which way such a confinement will influence the emergent
nonequilibrium micro-structure of the host medium.  On the one hand,
in~\cite{BCCMO01} the computation of the density profiles around a TP
in an adsorbed monolayer of hard-core particles showed that, if the
particle density in the monolayer is conserved, the approach to the
equilibrium density with the distance past the tracer is described by
a power-law function~\cite{footnote1}.  On the other hand, it is
known~\cite{BOMR96,OBBM} that in strictly one-dimensional systems --
the so-called single-files, there are no stationary profiles past the
driven tracer.  The systems we consider here are physically two-
(stripes) and three-dimensional (capillaries), but effectively they
are quasi-one-dimensional since they are of an infinite extent only in
one direction. Therefore, the outcome is \emph{a priori} not evident.
We show that the approach of the local host particles density at large
distances past the tracer is exponential, in sharp contrast with the
power-law behavior observed in unbounded systems~\cite{BCCMO01}. More
precisely, in the case of strip-like geometry of size $L$, our
analytical approach allows us to identify a characteristic length,
marking the passage from an initial algebraic decay at short distances
to an asymptotic exponential relaxation. This length scale diverges
with the size $L$ of the stripe, resulting in a pure algebraic decay
for infinite lattices, as expected. Our study reveals that the strong
memory effects in the spatial distribution of particles in the wake of
the tracer are suppressed by the confining geometry, due to a fast
homogenization effect.

The paper is organized as follows. In Section II we describe the model
and define the relevant parameters. In Section III we present our
results for $V(F)$ in the case of confined geometries and compare the
analytical predictions against Monte Carlo numerical simulations.  In
Section IV we study the phenomenon of NDM and discuss the effects of
different microscopic dynamical rules. In Section V we report the
analytical results for the density profiles around the tracer for a
strip-like geometry and compare them with the case of infinite
lattices. Finally, in Section VI we conclude with a brief
recapitulation of our results and outline further research. In
Appendices A and B, details on the analytical computations for the
force-velocity relation and the density profiles, respectively, are
provided.

\section{The model}

\begin{figure}[!t]
\includegraphics[width=1.\columnwidth,clip=true]{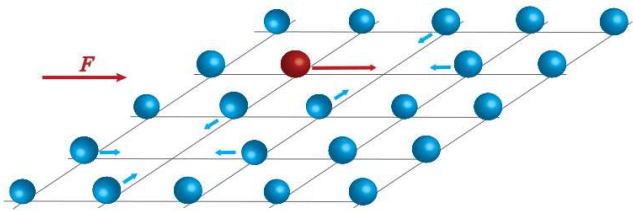}
\caption{Lattice gas with driven tracer. The hard-core tracer particle
  (in red) is subjected to an external force $F$ and performs a biased
  random walk, while the bath particles (in blue) perform
  nearest-neighbor lattice random walks, constrained by
  excluded-volume interactions.}
\label{fig0}
\end{figure}

\begin{figure}[!t]
\includegraphics[width=1.\columnwidth,clip=true]{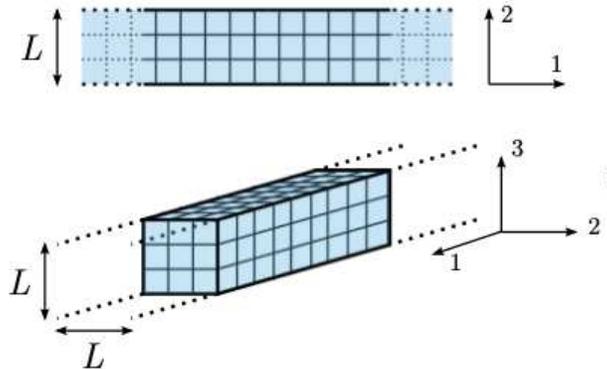}
\caption{Sketch of strip- and capillary-like confined geometries. The
  external force is applied along the direction in which the system has an infinite extent.}
\label{fig1}
\end{figure}

We consider a $d-$dimensional hyper-cubic lattice with unit lattice
spacing, bounded or not, populated by hard-core particles, with total
average density $\rho$ (see Fig.~\ref{fig0}).  The particle dynamics
is defined by the following rules: Each bath particle waits an
exponentially distributed time with mean $\tau^*$ and then selects one
of the nearest-neighboring sites with probability $1/2d$. If the
chosen site is empty at this time moment, the particle jumps;
otherwise, if the target site is occupied, the particle stays at its
position. We also introduce a tracer particle, with a different mean
waiting time $\tau$, and with jump probabilities in the direction
$\nu$ ($\nu\in \{\pm 1,\ldots,\pm d\}$)
\begin{equation}
p_\nu=\frac{e^{(\beta/2)\boldsymbol{F}\cdot\boldsymbol{e}_\nu}}
{\sum_\mu e^{(\beta/2)\boldsymbol{F}\cdot\boldsymbol{e}_\mu}},
\label{pnu}
\end{equation}
where $\beta$ is the inverse temperature (measured in the units of the
Boltzmann constant), $\boldsymbol{e}_\mu$ are the $2d$ base vectors of
the lattice, and $\boldsymbol{F}\equiv F\boldsymbol{e}_1$ is the
external force. In the following, we will refer to
  Eq.~(\ref{pnu}) as Choice 1.  Notice that in this case, in the limit
  $F\to\infty$, the TP performs a totally directed motion, where only
  the steps in the field direction are permitted.  This situation can
be relevant for modeling molecular motors, such as the protein kinesin
moving on microtubules~\cite{BBPZ15}.

Let us stress that the choice in Eq.~(\ref{pnu}) is not univocal. It
coincides with that of Ref.~\cite{LF13}, but the transition rates in
the transverse direction with respect to the field can have different
expressions, as considered for instance in~\cite{JKGC08,BM14}.  The
theoretical treatment proposed in the following is however general and
does not rely on a specific form of $p_\nu$. In order to illustrate
this point, and in order to analyze the effect of different forms of
jump probabilities, we will also explicitly consider the case of
Ref.~\cite{BM14}, which, in two dimensions, corresponds to the
definition
\begin{equation}
  p'_{\pm 1}=\frac{1}{2}\frac{e^{\pm \beta F/2}}
    {e^{\beta F/2}+e^{-\beta F/2}},
\label{maes}
\end{equation}
and $p'_{\pm 2}=1/4$. In this case, the jump probabilities in the
  directions orthogonal to the force are constant, and we will refer
  to this expression as Choice 2 in the rest of the paper. Note that
both Choice 1 and Choice 2 satisfy the LDB, which guarantees that, at
linear order in $F$, the Einstein relation is verified.  However, out
of equilibrium, or in the nonlinear regime, the explicit dependence of
the transition rates on $F$ can have important effects, as we will
discuss in detail in Section IV.

\section{Stationary velocity in confined geometries}

In this Section we present an analytical computation of the
stationary velocity of the tracer, based on a 
decoupling approximation.

\subsection{General formalism}

Let the Boolean variable $\eta(\boldsymbol{R})=\{1,0\}$ denote the
instantaneous occupation variable of the site at position
$\boldsymbol{R}$ by any of the bath particles, $\eta\equiv
\{\eta(\boldsymbol{R})\}$ denote the instantaneous configuration of
all such occupation variables and $\boldsymbol{R}_{TP}$ - the
instantaneous position of the TP.  The evolution of the joint
probability $P(\boldsymbol{R}_{TP},\eta; t)$ that at time $t$ the TP
is at site $\boldsymbol{R}_{TP}$ with the configuration of obstacles
$\eta$, is ruled by the master equation
\begin{eqnarray}
&&\partial_tP(\boldsymbol{R}_{TP},\eta; t)=\frac{1}{2d\tau^*} \nonumber \\
&\times&\sum_{\mu=1}^d
\sum_{\boldsymbol{r}\ne \boldsymbol{R}_{TP}-\boldsymbol{e}_{\mu},\boldsymbol{R}_{TP}}[P(\boldsymbol{R}_{TP},\eta^{\boldsymbol{r},\mu}; t)-P(\boldsymbol{R}_{TP},\eta; t)]
\nonumber \\
&+&\frac{1}{\tau}\sum_{\mu=1}^dp_\mu\{[1-\eta(\boldsymbol{R}_{TP})]P(\boldsymbol{R}_{TP}-\boldsymbol{e}_{\mu},\eta; t) \nonumber \\
&-&[1-\eta(\boldsymbol{R}_{TP}+\boldsymbol{e}_{\mu})]P(\boldsymbol{R}_{TP},\eta; t)\},
\end{eqnarray}
where $\eta^{\boldsymbol{r},\mu}$ is the configuration obtained from
$\eta$ by exchanging the occupation numbers of sites $\boldsymbol{r}$
and $\boldsymbol{r}+\boldsymbol{e}_\mu$. The first term in r.h.s. of
the above equation describes the change in
$P(\boldsymbol{R}_{TP},\eta; t)$ due to the bath particles jumps,
while the second and third lines take into account the TP moves.

Multiplying both sides of the master equation by $(\boldsymbol{R}_{TP}
\cdot \boldsymbol{e}_{1})$, summing over all configurations
$(\boldsymbol{R}_{TP},\eta)$, and taking the long time limit
$t\to\infty$, we obtain the following expression for the TP velocity
\begin{equation}
V = \frac{1}{\tau}\left\{p_1\left[1-k(\boldsymbol{e}_{1})\right]-p_{-1} \left[1-k(\boldsymbol{e}_{-1})\right]\right\},
\label{Vtp}
\end{equation}
where the function $k({\lamb})$ is the stationary value of
$k({\lamb};t)$, defined by
\begin{equation}
k({\lamb};t) =  \sum_{\boldsymbol{R}_{TP},\eta}
\eta(\boldsymbol{R}_{TP}+\lamb)P(\boldsymbol{R}_{TP},\eta;t).
\label{kk}
\end{equation}
The function $k({\lamb};t)$ represents the density profile around the
tracer position.  
The equation of motion for $k(\lamb;t)$ is obtained by
multiplying the master equation by $\eta(\boldsymbol{R}_{TP}+\lamb)$
and summing over all the configurations of
$(\boldsymbol{R}_{TP},\eta)$. We get the following equation:
\begin{eqnarray}
2d\tau^*\partial_t k(\lamb;t)&=&\sum_\mu\left(\nabla_\mu-\delta_{\lamb,\emuu}\nabla_{-\mu}\right)k(\lamb;t) \nonumber \\
&+&\frac{2d\tau^*}{\tau}\sum_\nu p_\nu \langle[1-\eta(\boldsymbol{R}_{TP}+\enu)]\nonumber\\
&\times&\nabla_\nu\eta(\boldsymbol{R}_{TP}+\lamb)\rangle,
\label{kl}
\end{eqnarray}
where we introduced the differential operator $\nabla_\mu
f(\lamb)=f(\lamb+\boldsymbol{e}_{\boldsymbol{\mu}})-f(\lamb)$ and the
average $\langle X(\boldsymbol{R}_{TP})\rangle\equiv
\sum_{\boldsymbol{R}_{TP},\eta}X(\boldsymbol{R}_{TP})P(\boldsymbol{R}_{TP},\eta;t)$.

Equation~(\ref{kl}) is not closed, because two-point correlation
functions of the occupation variable appear in the right hand side.
In order to close and solve this equation, we use the decoupling
approximation:
\begin{equation}
\langle \eta(\boldsymbol{R}_{TP}+\boldsymbol{\lambda})\eta(\boldsymbol{R}_{TP}+\boldsymbol{e}_\nu)\rangle \approx
\langle\eta(\boldsymbol{R}_{TP}+\boldsymbol{\lambda}) \rangle \langle \eta(\boldsymbol{R}_{TP}+\boldsymbol{e}_\nu)\rangle,
\label{approx}
\end{equation} 
which is expected to be valid for $\lamb\neq\boldsymbol{e}_\nu$.  
This approximation has been discussed in a series of previous papers
in unconfined geometries~\cite{BOMM92,BCLMO99,BCCMO00}. It is expected
to hold in the dilute regime, $\rho\ll 1$, and for values of $\tau^*$
not too large with respect to $\tau$, namely when the dynamics of bath
particles is sufficiently fast.  More specifically, it has been shown
that this approximation provides exact results for $V(F)$ in the
limits of very low and very high densities in unconfined
geometries~\cite{BIOSV14}. Notice that the regime of validity of the
approximation can be dependent on the specific choice of transition
rates.

\subsection{Solution in confined geometries}

The formalism developed up to here is general and holds for both
unconfined and confined lattices. The computation in the case of
infinite lattices has been reported in previous works, see
e.g.~\cite{BCCMO01}. Here we focus on the novel and interesting case
of confined geometries, namely infinite in the direction of the
applied field and finite of length $L$ in the other ones, with
periodic boundary conditions~\cite{footnote2}. In particular, for
$d=2$ we have stripes and for $d=3$ we have rectangular capillaries,
see Fig.~\ref{fig1}~\cite{footnote2bis}.  For convenience, let us
introduce the functions $h(\lamb;t)$, defined by
\begin{equation}
  h(\lamb;t)\equiv k(\lamb;t)-\rho,
\end{equation}
with the convention $h({\bf 0};t)=0$, and the shorthand notation
$h(n_1\boldsymbol{e}_1+\dots+n_d\boldsymbol{e}_d;t)=h_{n_1,\dots,n_d}(t)$.
As detailed in Appendix~\ref{stat_vel}, the function
$h_{n_1,\dots,n_d}$ introduced above is finally given by the following
system of $2d$ equations
\begin{eqnarray}
\alpha h_{n_1,\dots,n_d} &=& \sum_\nu A_\nu h_\nu \nabla_{-\nu} F_{n_1,\dots,n_d} \nonumber \\
&-& \rho (A_1 - A_{-1})(\nabla_1 -\nabla_{-1})F_{n_1,\dots,n_d}, \nonumber \\
\label{system}
\end{eqnarray}
where $(n_1,\dots, n_d)$ are taken equal to the coordinates of the
base vectors $\{ \pm \boldsymbol{e}_1,\dots, \pm \boldsymbol{e}_d\}$,
the coefficients
$A_\mu=1+(2d\tau^*/\tau)p_\mu[1-\rho-h(\boldsymbol{e}_{\mu})]$,
$\alpha = A_1+A_{-1}+2(d-1)A_2$, and we have introduced the functions
\begin{equation}
F_{\boldsymbol{n}} = \frac{1}{L^{d-1}} \sum_{k_2,\dots,k_d=0}^{L-1} \frac{1}{2\pi}\int_{-\pi}^\pi d q \frac{e^{-in_1q} \prod_{j=2}^d e^{-2i\pi n_j k_j/L}}{1- \lambda(q,k_2,\dots,k_d)},
\label{fn}
\end{equation}
with
\begin{equation}
\lambda(q,k_2,\dots,k_d) = \frac{A_1}{\alpha} \ex{-iq} + \frac{A_{-1}}{\alpha} \ex{iq}+\frac{2A_2}{\alpha}\sum_{j=2}^d \cos\left(\frac{2\pi k_j}{L} \right).
\end{equation}
Note that $F_{\boldsymbol{n}}$ is the long-time limit of the
generating function of a biased random walk on a strip-like
lattice~\cite{H95}.
Using the fact that $h_{\pm2}=\dots h_{\pm d}$ for symmetry reasons, this
system of $2d$ equations may be reduced to a system of three equations
for the coefficients $A_\nu$ ($\nu=\pm 1,2$). Although highly
nonlinear, the system~(\ref{system}) can be solved numerically and the
tracer velocity can be obtained for arbitrary values of the parameters
using the relation~(\ref{Vtp}), that can be rewritten as
\begin{equation}
V=\frac{1}{2 d \tau^*}\left (A_1 - A_{-1}\right).
\label{vstat}
\end{equation}

The system~(\ref{system}) and Eq.~(\ref{vstat}), with the
definition~(\ref{fn}), represent the first main result of our
paper. They allow us to obtain the prediction for $V(F)$ following
from the decoupling approximation~(\ref{approx}), for general
dimension $d$ and arbitrary values of the model parameters.

\subsection{Linearized solution in the dilute regime}

The system~(\ref{system}) can be simplified in the dilute limit,
$\rho\to 0$.  Noting that the coefficients $A_\nu$ can be approximated
as $A_\nu\sim 1+2dp_\nu\tau^*/\tau$, and introducing the variables
$q_{\bf n}$ through the relation
\begin{equation}
h_{n_1,\dots,n_d}=\rho\; q_{n_1,\dots,n_d},
\end{equation}
one can express the stationary TP velocity as
\begin{eqnarray}
V(\rho\to 0)&=& \frac{1}{\tau}(p_1-p_{-1}) \nonumber \\
&-&\frac{\rho}{\tau}(p_1-p_{-1}+p_1q_1-p_{-1}q_{-1})+o(\rho), \nonumber \\
\label{v_lin}
\end{eqnarray}
where now the quantities $q_{\bf n}$ satisfy the linear system
\begin{eqnarray}
2d\left(1+\frac{\tau^*}{\tau}\right)q_{\bf n}&=&\sum_\nu \left(1+2d\frac{\tau^*}{\tau}p_\nu\right) q_{{\bf e}_\nu}\nabla_{-\nu}F_{\boldsymbol{n}} \nonumber \\
&-&2d\frac{\tau^*}{\tau}(p_1-p_{-1})(\nabla_1-\nabla_{-1})F_{\boldsymbol{n}}. \nonumber \\
\label{sys_lin}
\end{eqnarray}
From Eqs.~(\ref{v_lin}) and~(\ref{sys_lin}), and using the
definition~(\ref{fn}), one can explicitly obtain the TP velocity in
the dilute limit, for arbitrary choice of jump probabilities and time
scales $\tau,\tau^*$. Notice that the comparison between the
expression for $V(F)$ obtained in the dilute limit for unconfined
geometries~\cite{BIOSV14}, following a computation analogous to that
reported here, and the analytical result of~\cite{LF13}, revealed that
the decoupling approximation is indeed exact at lowest order in
$\rho$, for arbitrary values of the time scales $\tau,\tau^*$. We
claim that this statement also holds for confined geometries.

\subsection{Comparison with numerical simulations}

\begin{figure*}[!t]
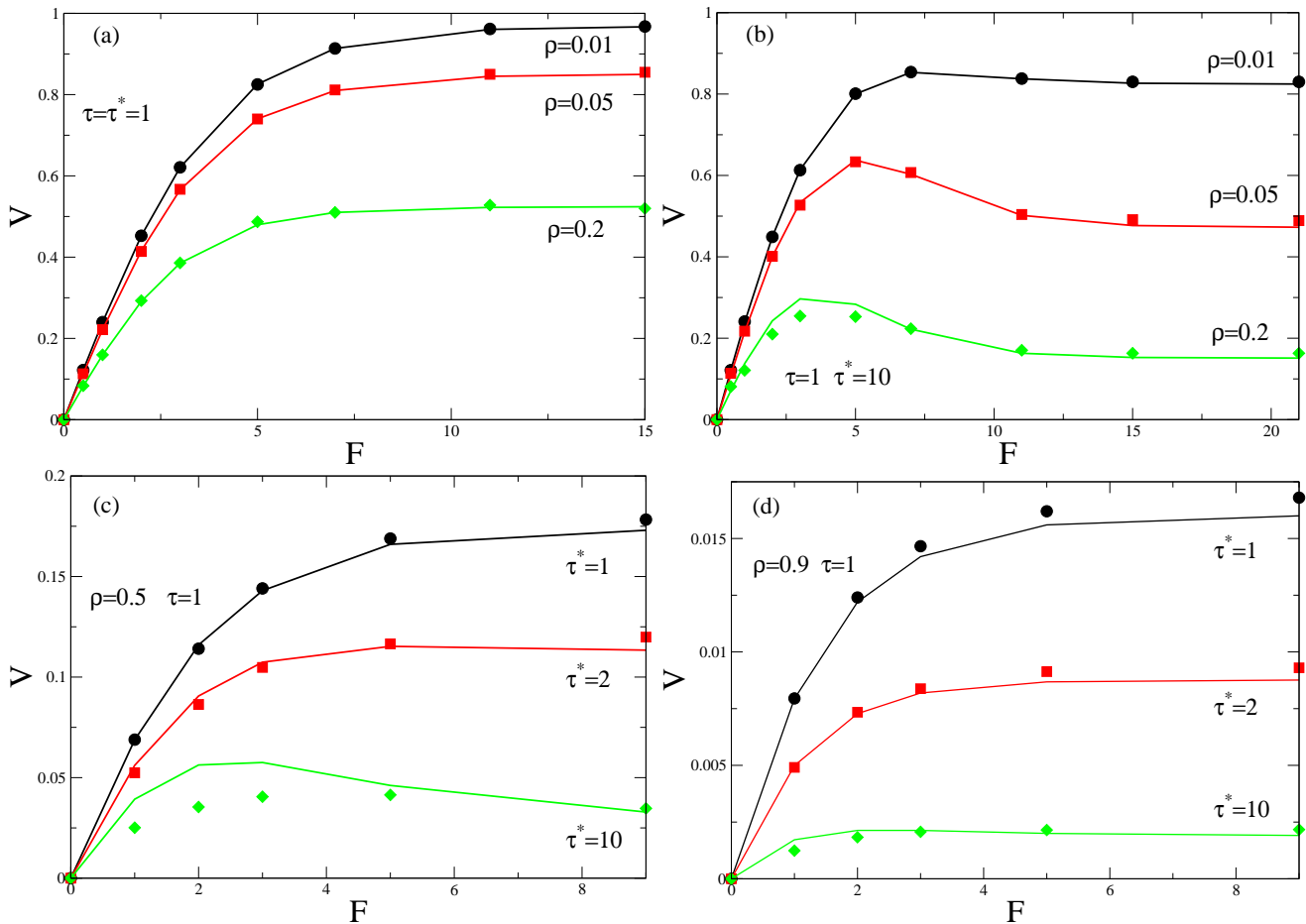

\includegraphics[width=1.\columnwidth,clip=true]{bande_taustar1.eps}
\includegraphics[width=1.\columnwidth,clip=true]{bande_taustar10.eps}
\includegraphics[width=1.\columnwidth,clip=true]{v_Frho05.eps}
\includegraphics[width=1.\columnwidth,clip=true]{v_F09.eps}
\caption{Analytical predictions (continuous lines) and numerical
  simulations (symbols) for $V(F)$ in the strip-like geometry with
  $L=3$, and jump probabilities of Choice 1. Parameters are:
  $\tau=\tau^*=1$ (a), $\tau=1, \tau^*=10$ (b), $\tau=1$ and
  $\rho=0.5$ (c), $\tau=1$ and $\rho=0.9$ (d).  Notice the
  nonmonotonic behavior in case (b).}
\label{fig2}
\end{figure*}

In order to check the validity of the approximation~(\ref{approx}) and
of the result~(\ref{vstat}), we have performed Monte Carlo numerical
simulations of our model.  We have considered a two-dimensional
strip-like lattice with $M=L_1\times L_2$ sites, with $L_1=6000$ and
$L_2=3$, and periodic boundary conditions in both directions. The
initial configuration is random with density $\rho=N/M$, $N$ being the
number of particles on the lattice.  In Fig.~\ref{fig2}, we report the
analytic results (continuous lines) obtained from the solution of the
system for the coefficients $A_\nu$, and the numerical data (symbols)
from Monte Carlo simulations for the force-velocity relation, for
different values of density and time scales, in the case of
  Choice 1. A very good agreement is observed in a wide range of
parameters. As expected, at finite densities, the decoupling
approximation turns out to be less accurate at large values of
$\tau^*$.  We also notice that in the case of high density $\rho=0.9$,
we found that the relaxation times of the system become very long and
an accurate numerical estimation of the stationary velocity requires
long simulations.

\section{Negative differential mobility and role of microscopic dynamical rules}

As shown in Fig.~\ref{fig2}(b), in the case with $\tau=1$
and $\tau^*=10$, the stationary velocity of the TP shows NDM, namely a
nonmonotonic behavior with the applied force.  In the framework of
lattice gases~\cite{LF13,BM14,BIOSV14} this behavior occurs due to the
formation of long-lived traps caused by the crowding of bath particles
in front of the tracer. Since the TP cannot push the obstacles away,
when the dynamics of bath particles is slow enough, namely for large
$\tau^*$, an increasing of the external force produces a longer escape
time from the traps, resulting in a reduction of the average drift.

\begin{figure*}[!t]
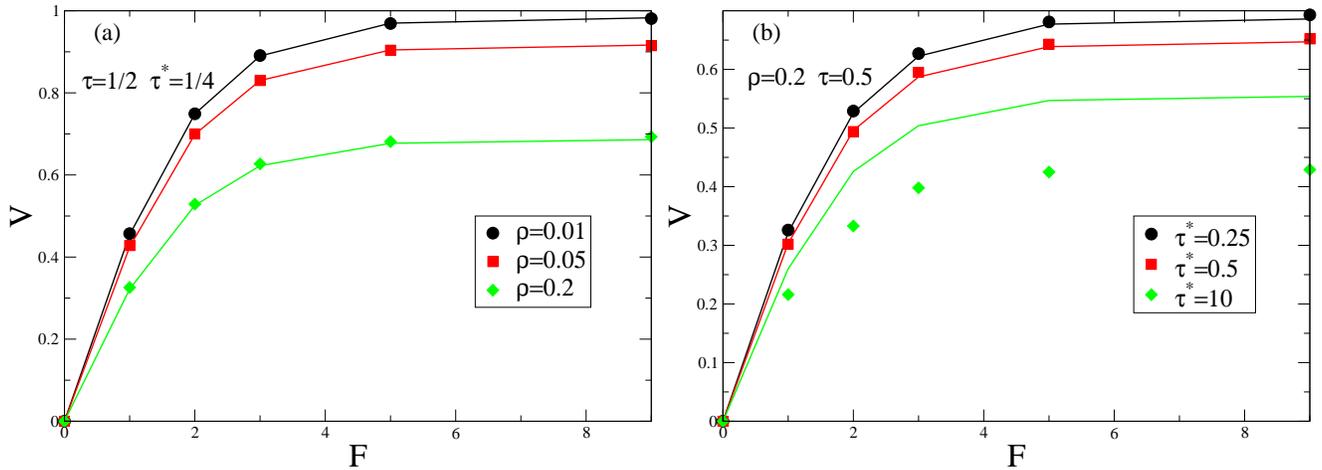

\includegraphics[width=1.\columnwidth,clip=true]{v_F_maes.eps}
\includegraphics[width=1.\columnwidth,clip=true]{V_F_02Ts.eps}
\caption{Analytical predictions (continuous lines) and numerical
  simulations (symbols) for $V(F)$ in the strip-like geometry with
  $L=3$, and with jump probabilities of Choice 2, with $\tau=0.5$ and
  $\tau^*=0.25$, for different values of $\rho$ (a), and $\rho=0.2,
  \tau=0.5$ and different values of $\tau^*$ (b).}
\label{fig_maes}
\end{figure*}

\begin{figure}[!t]
\includegraphics[width=1.\columnwidth,clip=true]{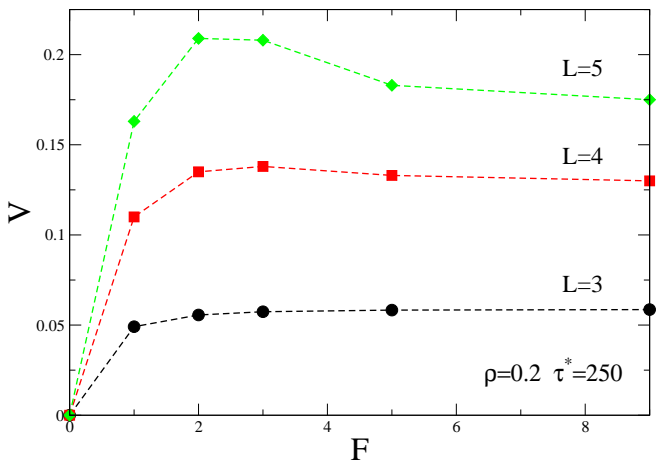}
\caption{Numerical simulations for $V(F)$ in the strip-like geometry
  with $L=3,4,5$ for the model with transition probabilities of Choice
  2, with $\tau=0.5$, $\tau^*=250$ and $\rho=0.2$ (dashed lines are a
  guide to the eye). The nonmonotonic behavior can be observed for
  $L\ge 4$.}
\label{fig_maes2}
\end{figure}

The complete phase chart identifying the region in the parameter space
$\{\rho,\tau^*/\tau\}$ where the phenomenon is expected to occur has
been obtained in~\cite{BIOSV14} for infinite lattices, in the
  case of Choice 1. Here we focus on the case of confined geometry
and on the role of the microscopic dynamical rules. Indeed, these can
deeply modify the region in the parameter space where NDM occurs.  In
order to elucidate this important point, we also apply our theory to
the model where the transition rates are independent of the field in
the transverse direction. These rates have been used for instance in
the numerical simulations described in~\cite{BM14}, and, in our
formalism, correspond to Choice 2.  As mentioned above, in our
theory the specific form of transition rates only enters the last step
of the computation, where the numerical solution of the nonlinear
system for the coefficients $A_\nu$, Eq.~(\ref{system}), is carried
out. Therefore, the application to the model with jump probabilities
of Choice 2 is straightforward.  In Fig.~\ref{fig_maes} we compare the
prediction of our analytic solutions for this model with the results
of numerical simulations for some cases, finding a very good agreement
for not too large values of $\tau^*$.

As shown numerically in~\cite{BM14} for unconfined geometries, and as
also discussed in~\cite{BIOSV14}, a nonmonotonic behavior of $V(F)$
for Choice 2 is observed for large values of
$\tau^*/\tau$~\cite{footnote3}. In this regime, the decoupling
approximation is expected not to be accurate, because the bath
particles are very slow and their motion is strongly correlated.
Moreover, it is important to notice that, in the case of Choice 2, in two
dimensions, trapping arises as a three-particle effect: indeed, in
order to form a trap one needs an obstacle in front of the TP and two
others on the adjacent sites, in the orthogonal directions (see also
the discussion in~\cite{BSV15}).  Therefore, in the case of confined
geometries, a NDM cannot be observed for a stripe with $L=3$, because
in this case the trapping time becomes independent of the applied
force (the TP has no chance to escape the trap and has to wait for the
obstacles to step away, for any value of the external field). However,
interestingly, as shown in Fig.~\ref{fig_maes2} via numerical
simulations, the phenomenon occurs for $L\ge 4$, and sufficiently
large values of $\tau^*/\tau$.

\begin{figure*}[!t]
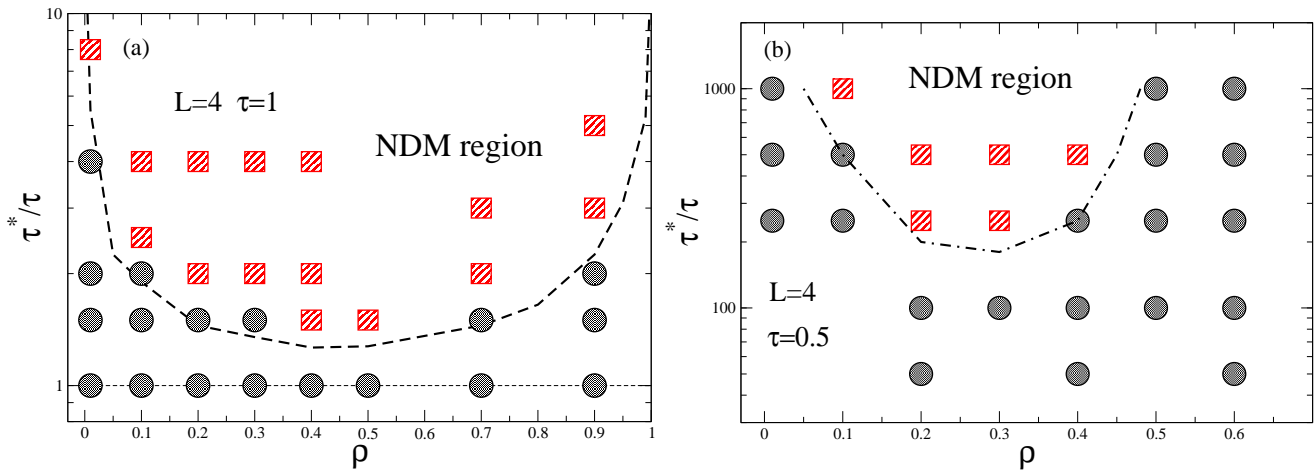

\includegraphics[width=1.\columnwidth,clip=true]{phase_chart.eps}
\includegraphics[width=1.\columnwidth,clip=true]{phase_chart_maes2.eps}
\caption{(a) Phase chart for the NDM region in the plane $\tau^*/\tau$
  vs $\rho$ in the case of Choice 1, for a strip-like geometry with
  $L=4$.  The bold dashed line represents the line separating the
  region where NDM occurs, according to the analytical solution based
  on the decoupling approximation. (b) Phase chart for the model with
  jump probabilities of Choice 2. The dot-dashed line is a guide to
  the eye. The red squares (black circles) represent numerical
  simulations where the NDM effect is (is not) observed.}
\label{fig_fc}
\end{figure*}

In order to obtain a complete picture and to explicitly show the
qualitative difference of behavior that the choice of transition rates
can produce, we have reconstructed the phase charts where the effect
of NDM is expected to occur, in the parameter space $\tau^*/\tau$ vs
$\rho$, for both Choice 1 and Choice 2, focusing on the case of
a strip-like geometry with $L=4$, see Fig.~\ref{fig_fc}.  In the case
of Choice 1, the decoupling approximation turns out to be very
accurate in predicting quantitatively the range of parameters for
which the NDM occurs, as represented by the bold dashed line in
Fig.~\ref{fig_fc}(a). Moreover, in this case, from the linearized
solution, Eqs.~(\ref{v_lin}) and~(\ref{sys_lin}), an exact criterion
for NDM can be obtained in the low density limit. Following the
procedure described in~\cite{BIOSV14}, we find that the line
separating the region of NDM from that where the effect does not occur
in the plane $\tau^*/\tau$ vs $\rho$ obeys the same scaling observed
in unconfined geometries, namely $\tau^*/\tau\sim 1/\sqrt{4\rho}$, for
$\tau^*/\tau\to\infty$. 

In the case with jump probabilities of Choice 2, it turns out
  that the analytical approach based on the decoupling approximation
  predicts NDM in a region of the parameter space where it is actually
  not observed via numerical simulations.  This is due to the larger
  values of $\tau^*/\tau$ required for NDM to occur, so that the
  decoupling approximation is not accurate, and cannot account
  quantitatively for NDM. The phase chart for this choice of
  transition rates obtained via numerical simulations is reported in
  Fig.~\ref{fig_fc}(b)~\cite{footnote3bis}.

Our results bring to the fore the important and general issue
concerning the choice of microscopic dynamical rules for
nonequilibrium systems. Indeed, in the nonlinear regime, different
transition rates, all satisfying the LDB, can produce different
behaviors. The correct choice should be dictated by physical
considerations, and depends on the real system under study. On the one
hand, Choice 1 includes the case of totally directed motion, that can
be useful to model the dynamics of molecular motors, or the case of a
force which reduces the excursions of the TP in the transverse
directions. On the other hand, Eq.~(\ref{maes}) can be more realistic
for systems where the motion in the different directions is totally
decoupled, as, e.g., for Brownian dynamics in two or three dimensions,
with a force applied along a given direction.

To sum up: i) the
  general formalism proposed holds for any form of jump probabilities,
  in particular for Choice 1 and Choice 2; ii) NDM is observed in both
  cases in confined geometries, as soon as $\L\ge 4$ for the case of
  Choice 2; iii) our analytical results based on the decoupling
  approximation account quantitatively for the phenomenon in the case
  of Choice 1; iv) the theory predicts only qualitatively NDM in the
  case of Choice 2.

\section{Density profiles around the tracer particle}

In this Section we focus on the study of the bath particle density
distribution around the tracer. This quantity plays a central role in
the characterization of the micro-structural changes induced by the
probe in the host medium and these heterogeneous deformations are
responsible for the effective drag coefficient acting on the tracer.

\subsection{Two-dimensional infinite lattice}

The density profile at large distance from the probe for the case of a
two-dimensional infinite lattice has been described
in~\cite{BCCMO01}. It is found that the decay of the density profile in front and past
the TP shows two different behaviors. In particular, the asymptotic
large distance behavior for the density profile in front of the tracer is exponential:
\begin{equation}
h_{n,0}\sim K_+\frac{e^{-n/\lambda_+}}{n^{1/2}},
\label{h_inf}
\end{equation}
where $K_+$ is a coefficient depending on $\{A_\nu\}$ and $\rho$ (the
explicit expression is given in~\cite{BCCMO01}), and $\lambda_+$ is
the decay length, see~\cite{BCCMO01}. Remarkably, the
wake past the tracer particle shows a different, algebraic, behavior
\begin{equation}
h_{-n,0}=- \frac{K_-}{n^{3/2}}\left( 1+\frac{3}{8n}+
\mathcal{O}(1/n^2)\right),
\label{h_behind_inf}
\end{equation}
where the coefficient $K_-$ depends on $\{A_\nu\}$ and $\rho$, and is
reported in~\cite{BCCMO01}. Note that this algebraic decay has also
been shown to occur in soft dense colloids~\cite{DBJ14}.

\subsection{Strip-like geometry}

In confined geometries the theoretical prediction of the density
profile around the TP can be obtained numerically for arbitrary values
of the distance $n$ from Eq.~(\ref{system}).  An explicit analytical
form can be derived for large distances from the TP.  In
Appendix~\ref{den_prof} we present the detailed computation of the
asymptotic density profile around the tracer focusing on the case of a
strip-like geometry, namely for a two-dimensional lattice, infinite in
the field direction and of size $L$ in the transverse direction, with
periodic boundary conditions. The main result of our computation is
that, at variance with the unconfined case, the density profile
displays an exponential decay both in front \emph{and} past the
tracer.

\subsubsection{Density profiles in the wake of the tracer}
 
As shown in Appendix~\ref{den_prof}, 
the density profiles in the wake of the tracer in the
limit $n \to -\infty$ is
 \begin{equation}
h_{n,0} \underset{n\to-\infty}{\sim} \mathcal{H}^- \ex{n/\ell^-},
\label{h_behind}
\end{equation}
with
\begin{eqnarray}
\mathcal{H}^-  & = & \frac{2}{L\At_1[U_1(1)-U_2(1)]}  \Big\{ \At_1 h_1 [U_1(1)^{-1} -1] \nonumber \\
&+&  \At_{-1} h_{-1} [U_1(1) -1] +2\At_2 h_2 \left( \cos \frac{2\pi}{L}-1\right) \nonumber \\
&-& \rho(\At_1-\At_{-1}) [U_1(1)-U_1(1)^{-1}] \Big\}, \\
\ell^- & = & \frac{1}{\ln [U_1(1)]},
\label{Hm}
\end{eqnarray}
where $\At_\mu$ and the function $U_1$ are defined in
Appendix~\ref{den_prof}.  One can check that $U_1(1)>1$ so that
$\ell^->0$.  Our computation shows that, in contrast with what was
found for infinite lattices, the density profile past the tracer in
confined geometries presents an exponential decay. This is a
surprising result, because one could have expected a behavior
intermediate between that of an infinite lattice and that of a
one-dimensional single-file for which there is no stationary density
profile~\cite{BOMR96}.  On the contrary, what we find here is
analogous to the behavior observed in the study of biased diffusion in
a one-dimensional adsorbed monolayer~\cite{BCLMO99}. The exponential
decay is related to the fast homogenization in the transverse
direction due to confinement. Let us also notice that from the
solution of the system~(\ref{system}) and the expression~(\ref{Hm}),
we obtain that the decay length $\ell^-$ diverges with the stripe width
$L$.
 
\begin{figure}[!t]
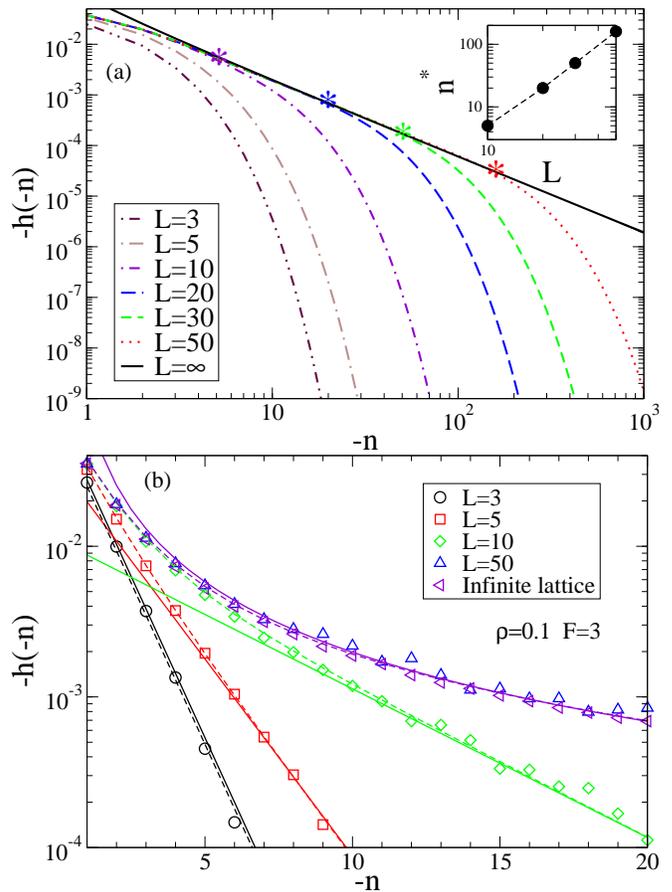

\includegraphics[width=1.\columnwidth,clip=true]{profilo_L.eps}
\includegraphics[width=1.\columnwidth,clip=true]{profilo_varieL.eps}
\caption{(a) Analytical predictions from Eq.~(\ref{system}) for the
  (rescaled) density profile past the tracer $h(n)$, for different
  values of the stripe length $L=3,5,10,20,30,50$, with parameters
  $\tau=\tau^*=1$, density $\rho=0.1$, and $F=3$. The black continuous
  line represents the asymptotic prediction in the case of an infinite
  lattice, Eq.~(\ref{h_behind_inf}). The stars represent the estimated
  values of the crossover length $n^*$.  Inset: characteristic
  crossover length $n^*$ as a function of $L$. (b) Analytical
  predictions from Eq.~(\ref{system}) (dashed lines), asymptotic
  analytic predictions, Eqs.~(\ref{h_behind}-\ref{Hm})
  and~(\ref{h_behind_inf}) (continuous lines), and numerical
  simulations (symbols) of $h(n)$, for different values of the strip
  length $L=3,5,10,50$, with $\tau=\tau^*=1$.}
\label{fig6bis}
\end{figure}

In Fig.~\ref{fig6bis}(a) we show the (rescaled) density profile $h(n)$
past the tracer obtained from the complete analytic solution of
Eq.~(\ref{system}), for different stripe lengths $L$. Interestingly,
we observe a change of behavior from an initial algebraic relaxation,
to a final exponential decay.  This allows us to identify a
characteristic length scale $n^*$, governing the crossover between the
two regimes. This length diverges with increasing $L$, as expected
from the known results for infinite lattices. In particular, it is
found that the behavior of the crossover length as a function of $L$
follows a power law, $n^*(L)\sim L^\gamma$, with $\gamma \simeq 2$,
see inset of Fig.~\ref{fig6bis}(a). This can be understood by
considering that the diffusional time $t^*$ for homogenization in the
transverse direction is $t^*\sim L^2/D$, where $D$ is a diffusion
coefficient. Since the TP travels at average velocity $V$, the
characteristic crossover length can be then estimated as $n^*\sim V
t^* \sim V L^2/D$.

\begin{figure*}[!t]
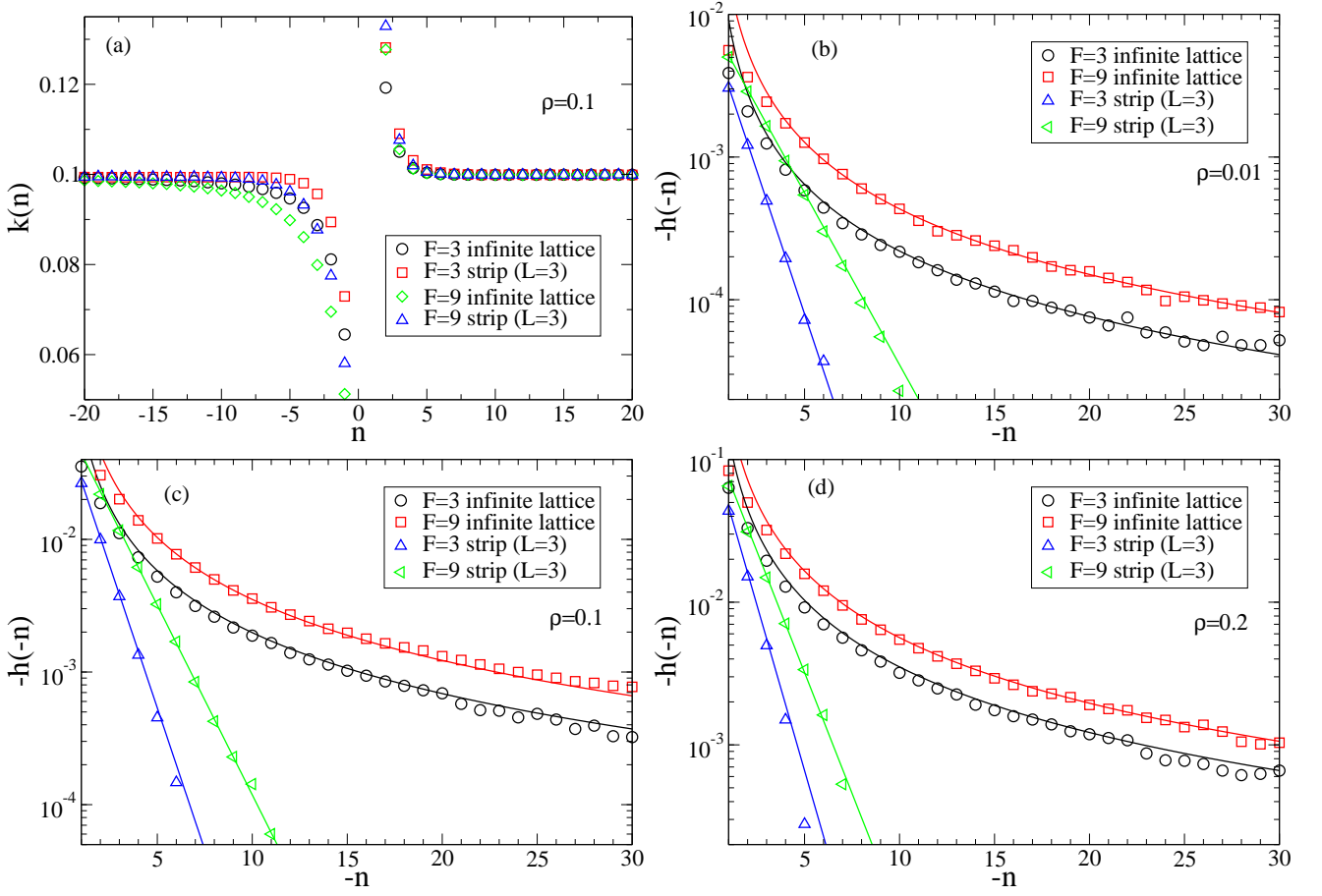

\includegraphics[width=1.\columnwidth,clip=true]{profile_sim01.eps}
\includegraphics[width=1.\columnwidth,clip=true]{profile_001.eps}
\includegraphics[width=1.\columnwidth,clip=true]{profile_01.eps}
\includegraphics[width=1.\columnwidth,clip=true]{profile_02.eps}
\caption{(a) Numerical simulations results for the stationary density
  profile $k(n)$ (defined in Eq.~(\ref{kk})) in the infinite lattice
  and in the strip-like geometry, with $\tau=\tau^*=1$, $\rho=0.1$
  and different values of the force $F$. (b)-(d) Asymptotic analytical
  predictions, Eqs.~(\ref{h_behind}-\ref{Hm}) and~(\ref{h_behind_inf})
  (continuous lines) and numerical simulations (symbols) for the
  (rescaled) density profile past the tracer $h(n)$ in the infinite
  lattice and in the strip-like geometry, with $\tau=\tau^*=1$ and
  different values of $\rho$.}
\label{fig4}
\end{figure*}
    
\subsubsection{Density profiles in front of the tracer}

In Appendix~\ref{den_prof} we show that, considering for simplicity
the special case where $L$ is even, the density profiles in front of
the tracer at large distances are given by
\begin{equation}
h_{n,0} \underset{n\to\infty}{\sim} \mathcal{H}^+ \ex{-n/\ell^+},
\label{h_front}
\end{equation}
with
\begin{eqnarray}
\mathcal{H}^+  & = & \frac{2}{L\At_1[U_1(L/2)-U_2(L/2)]}  \Big\{ \At_1 h_1 [U_2(L/2)^{-1} -1] \nonumber \\
&+&  \At_{-1} h_{-1} [U_2(L/2) -1] +2\At_2 h_2 \left( \cos \frac{2\pi}{L}-1\right) \nonumber \\
&&- \rho(\At_1-\At_{-1}) [U_2(L/2)-U_2(L/2)^{-1}] \Big\}, \\
\ell^+ & = &-\frac{1}{\ln [U_2(L/2)]},
\label{Hp}
\end{eqnarray}
where the function $U_2$ is defined in Appendix~\ref{den_prof}.  One
can check that $U_2(L/2)<1$ so that $\ell^+>0$. The characteristic
length $\ell^+$ rapidly converges to the value of the unconfined
geometries with the stripe size $L$.

\subsection{Comparison with numerical simulations}

In Fig.~\ref{fig6bis}(b), we compare results of numerical simulations and
analytical predictions for different values of the stripe length $L$,
at fixed density $\rho=0.1$ and fixed characteristic times
$\tau=\tau^*=1$, finding a very good agreement. The continuous lines
represent the asymptotic results, Eqs.~(\ref{h_behind}-\ref{Hm})
and~(\ref{h_behind_inf}), while the dashed lines represent the
complete solution for $h(n)$, valid for arbitrary $n$, obtained from
Eq.~(\ref{system}). Notice that for intermediate values of $L$ a
crossover between the algebraic and the exponential decay can be
observed, as predicted by our analytical solution.

The density profiles around the tracer $k(n)$ measured in numerical
simulations for density $\rho=0.1$ and two values of the force
$F=3,9$, are reported in Fig.~\ref{fig4}(a), in the 
case of Choice 1~\cite{footnote4}.  In Figs.~\ref{fig4}(b)-(d) we report the
(rescaled) bath particle density profiles $h(n)$ in the wake past the
TP in both infinite and confined geometries, for different values of
$\rho$ and $F$. The analytical asymptotic predictions for infinite
lattices, Eq.~(\ref{h_behind_inf}), and confined geometries,
Eqs.~(\ref{h_behind}-\ref{Hm}), represented by continuous lines, are
in very good agreement with the numerical simulations (symbols).  In
simulations the infinite lattice is realized with $L_1=L_2=200$, with
periodic boundary conditions, and the strip-like geometry with
$L_1=6000$ and $L_2=3$.

\section{Conclusion}

We have studied analytically the dynamics of a driven tracer,
traveling in a sea of unbiased hard-core particles on a lattice, in
the relevant case of confined geometries. Our general theory unifies
and extends previous results, can be applied to any choice of jump
probabilities and for arbitrary particle density, and allows us to
obtain the force-velocity relation of the TP and the density profiles
around it.  The central point of our approach is based on a decoupling
approximation for the site occupation variable correlation function.

Our treatment allowed us to address some important issues in the study
of the nonlinear response of a many-body interacting system.  First,
regarding the force-velocity relation of the TP, we have shown that
our analytical theory is in very good agreement with Monte Carlo
numerical simulations, in a wide range of model parameters. In
particular, we have found that, in confined geometries also, $V(F)$ in
the nonlinear regime can display a nonmonotonic behavior, namely a
negative differential mobility. This phenomenon is due to the trapping
effect induced by the bath particle on the tracer when the applied
field is strong. If the mean characteristic time of the bath particles
is sufficiently large, the traps can be long-lived and this produces a
decrease of the stationary drift as a function of the force.  The
microscopic mechanism inducing the trapping effect can depend on the
specific dynamical rules of the model~\cite{BIOSV14,BSV15}.  We have
discussed in detail the different behaviors arising from two
particular definitions of jump probabilities~\cite{BM14,BIOSV14},
identifying the region in the parameter space where NDM can take
place. In the case of transition rates corresponding to the case of
Choice 1, our theory successfully predicts the phase chart where NDM
is expected to occur. In the case of Choice 2, the decoupling
approximation qualitatively predicts NDM, but cannot account for the
phenomenon in the right region of the parameter space, because NDM
takes place only when the bath particles are very slow, and their
motion is strongly correlated.

The second important issue we have addressed regards the study of the
density profiles around the TP, and the local deformations induced in
the surrounding medium.  We have found that, quite surprisingly, the
density profile past the tracer strongly depends on the geometry of
the system. In the case of an infinite lattice with $d>1$, it is known
that the decay is algebraic~\cite{BCCMO01}, while for strip-like
geometries our theory predicts an exponential decay.  Our analytical
results also allow us to describe the passage from the algebraic decay
to the exponential relaxation, as a function of the strip
size. Remarkably, we found that this change of behavior is governed by
a characteristic length arising in the density profile, that
  scales with the stripe size as $\sim L^2$. For short distances with
respect to this length scale an algebraic decay is expected, while at
larger distances the exponential relaxation takes place. Since this
length scale diverges with the stripe size, for infinite lattices only
the algebraic decay survives.

Our findings show how the structure and shape of the density profile
are significantly affected by the lattice geometry. This can have
important implications for many known phenomena related to the
perturbation profile induced by a tracer in a host medium, such as
drifting steady-state structures~\cite{LZ97}, bath-mediated
interactions of driven particles~\cite{MO10,KL15}, and ``negative''
mass transport~\cite{LK10}. Moreover, it would be interesting to check
if a change of behavior similar to that observed for discrete lattices
between infinite and confined geometries is also found in off-lattice
systems, such as colloidal suspensions or granular media.

\section*{Acknowledgments}
The authors wish to thank C. Maes, U. Basu and M. Baiesi for helpful
discussions.  OB and AS acknowledge financial support from the ERC
Grant No. FPTOpt-277998.

\appendix
\section{Stationary velocity}
\label{stat_vel}

In this Appendix we present the details on the computations of the TP stationary velocity
in a confined geometry in dimension $d$.
The functions $h(\lamb;t)$, defined by
\begin{equation}
h(\lamb;t)\equiv k(\lamb;t)-\rho,
\end{equation}
with the convention $h({\bf 0};t)=0$,  satisfy the following equations of motion
\begin{eqnarray}
\label{hbulk}
2d\tau^*\partial_t h(\lamb;t) &=& \widetilde{L} h(\lamb;t) \quad \textrm{for} \quad {\boldsymbol \lambda} \notin \{{\bf 0},\pm\boldsymbol{e}_{1},\ldots,\pm\boldsymbol{e}_{d}\} \nonumber  \\
&& \\
2d\tau^*\partial_t h(\lamb;t) &=& \widetilde{L} h(\lamb;t) + \rho(A_\nu - A_{-\nu}) \nonumber \\
&& \quad \textrm{for} \quad {\boldsymbol \lambda} \in \{{\bf 0},\pm\boldsymbol{e}_{1},\ldots,\pm\boldsymbol{e}_{d}\},
\label{hboundary}
\end{eqnarray}
with $\widetilde{L}\equiv\sum_\mu A_\mu\nabla_\mu$ and
$A_\mu=1+(2d\tau^*/\tau)p_\mu[1-\rho-h(\boldsymbol{e}_{\mu})]$. 

We introduce the auxiliary variable
$\boldsymbol{\xi}=(\xi_1,\dots,\xi_d)$ and the generating function
\begin{equation}
H(\boldsymbol{\xi};t) = \sum_{n_1=-\infty}^\infty \sum_{n_2,\cdots,n_d=0}^{L-1} h_{n_1,\cdots,n_d}(t)\prod_{j=1}^d\xi_j^{n_j},
\end{equation}
where the shorthand notation
$h(n_1\boldsymbol{e}_1+\dots+n_d\boldsymbol{e}_d;t)=h_{n_1,\dots,n_d}(t)$
has been used. If $(n_1,\dots,n_d)=\boldsymbol{e}_\nu$ then we denote
$h_{\boldsymbol{e}_\nu}=h_\nu$.
From Eqs.~(\ref{hbulk}) and~(\ref{hboundary}) we can show that
$H(\boldsymbol{\xi};t)$ is the solution of the following partial
differential equation
\begin{eqnarray}
2d\tau^* \partial_t H(\boldsymbol{\xi};t) &=& 
\left[ \frac{A_1}{\xi_1}+A_{-1}\xi_1 + A_2\sum_{j=2}^d\left(\frac{1}{\xi_j} + \xi_j \right)  - \alpha \right]\nonumber \\
&\times& H(\boldsymbol{\xi};t)+K(\boldsymbol{\xi};t),
\label{H}
\end{eqnarray}
with $\alpha = A_1+A_{-1}+2(d-1)A_2$ and
\begin{eqnarray}
\label{defK}
K(\boldsymbol{\xi};t)&=&A_1(\xi_1-1)h_{1}(t)+A_{-1}\left(\frac{1}{\xi_1}-1\right)h_{-1}(t)\nonumber\\
&+&A_2 \sum_{j=2}^d\left[ (\xi_j-1)h_j(t) +\left(\frac{1}{\xi_j}-1\right)h_{-j}(t) \right]\nonumber \\
&+& \rho(A_1-A_{-1}) \left( \xi_1-\frac{1}{\xi_1}\right).
\end{eqnarray}
The stationary solution of Eq.~(\ref{H}) is
\begin{equation}
H(\boldsymbol{\xi})=\frac{K(\boldsymbol{\xi})}{\alpha} \frac{1}{1-\left[\frac{A_1}{\alpha} \frac{1}{\xi_1} + \frac{A_{-1}}{\alpha}\xi_1+\frac{A_2}{\alpha} \sum_{j=2}^d\left(\frac{1}{\xi_j}+\xi_j \right)  \right]}.
\end{equation}
We then rewrite the auxiliary variables as $\xi_1 = \ex{iq}$ and
$\xi_j = \ex{2i\pi k_j/L}$ (for $j\neq 1$), and introduce the function
\begin{equation}
F_{\boldsymbol{n}} = \frac{1}{L^{d-1}} \sum_{k_2,\dots,k_d=0}^{L-1} \frac{1}{2\pi}\int_{-\pi}^\pi d q \frac{e^{-in_1q} \prod_{j=2}^d e^{-2i\pi n_j k_j/L}}{1- \lambda(q,k_2,\dots,k_d)},
\label{fn2}
\end{equation}
with
\begin{equation}
\lambda(q,k_2,\dots,k_d) = \frac{A_1}{\alpha} \ex{-iq} + \frac{A_{-1}}{\alpha} \ex{iq}+\frac{2A_2}{\alpha}\sum_{j=2}^d \cos\left(\frac{2\pi k_j}{L} \right),
\end{equation}
so that $H(\boldsymbol{\xi})$ can be cast in the form
\begin{equation}
H(q,k_2,\dots,k_d) = \frac{K(q,k_2,\dots,k_d)}{\alpha}\frac{1}{1-\lambda(q,k_2,\dots,k_d)}.
\end{equation}
Using the definition of $F_{\boldsymbol{n}}$, we
get
\begin{eqnarray}
\frac{1}{1-\lambda(q,k_2,\dots,k_d)} &=& \sum_{n_1=-\infty}^\infty \sum_{n_2,\dots,n_d=0}^{L-1} e^{in_1q}\nonumber \\
&\times& \prod_{j=2}^d e^{2i\pi n_j k_j/L} F_{n_1,\dots,n_d}, \nonumber \\
\end{eqnarray}
and
\begin{eqnarray}
H(q,k_2,\dots,k_d) &=& \frac{1}{\alpha}  \sum_{n_1=-\infty}^\infty \sum_{n_2,\dots,n_d=0}^{L-1}  K(q,k_2,\dots,k_d) \nonumber \\
&\times&F_{n_1,\dots,n_d} e^{in_1q} \prod_{j=2}^de^{2i\pi n_j k_j/L}.
\end{eqnarray}
Finally, the definition of $K$ in Eq.~(\ref{defK}) allows us to show
that $h_{n_1,\dots,n_d}$ is given by the system of $2d$
equations~(\ref{system}).

\section{Density profiles}
\label{den_prof}

In this Appendix we present details on the computation of the density profiles
around the TP in the case of a strip-like geometry.
For simplicity, we introduce $\widetilde{A}_\nu = A_\nu /\alpha$ (and
then $\sum_{\nu} \widetilde{A}_\nu=1$).  Note that Eq.~(\ref{system})
is not closed with respect to the density profiles
$h_{\boldsymbol{n}}$. A closed form can be obtained as follows:
evaluating Eq.~(\ref{system}) for
$\boldsymbol{n}=\ee_1,\ee_{-1},\ee_2$, one obtains a closed set of
equations satisfied by $h_1$, $h_{-1}$ and $h_2$. These boundary
values of $h_{\boldsymbol{n}}$ can be obtained, and
$h_{\boldsymbol{n}}$ for arbitrary $\boldsymbol{n}$ can be computed
using Eq. (\ref{system}).

In order to obtain an explicit expression for the density profiles, in
what follows, we study the asymptotic behavior of $h_{n,0}$ in the
limits $n \to \pm \infty$.  To this aim, we consider the behavior of
the gradients $\nabla_\nu F_{n,0}$ in the limits $n \to \pm
\infty$. We will use the following expression of $F_{\boldsymbol{n}}$,
that can be obtained with the change of variable $u=\ex{\ii q}$ and by
computing the integral over $u$ with the residue theorem:
 
\begin{eqnarray}
  F_{\boldsymbol{n}} &=& \frac{1}{L} \sum_{k}^{L-1} \frac{1}{2\pi}\int_{-\pi}^\pi \dd q \frac{\ex{-\ii n_1 q}\ex{-2\ii\pi n_2 k/L}}{1- \lambda(q,k)}\\
  & = & \frac{1}{L} \sum_{k=0}^{L-1} \ex{-\frac{2\ii \pi k n_2}{L}} f(n_1,k),
 \label{rewriteF}
\end{eqnarray}
 with
 \begin{equation}
\label{deff}
f(n_1,k) = \begin{dcases}
  \frac{1}{\widetilde{A}_1} \frac{{U_2(k)}^{n_1}}{U_1(k)-U_2(k)}  & \text{if $n_1 \geq 0$ } \\
  \frac{1}{\widetilde{A}_1} \frac{{U_1(k)}^{n_1}}{U_1(k)-U_2(k)}   & \text{if $n_1 < 0$ },
	\end{dcases}
\end{equation}
 and
 \begin{eqnarray}
\label{defU}
U_{\substack{1 \\ 2}}(k) &=& \frac{1}{2\At_1 }\left( 1-2\At_2\cos \frac{2\pi k}{L} \right) \nonumber \\
&\pm& \frac{1}{2}\sqrt{\frac{1}{{\At_1}^2}\left(  1-2{\At_2} \cos \frac{2\pi k}{L}   \right)^2-4\frac{\At_{-1}}{\At_{1}}}.
\end{eqnarray}
 
We first study the behavior of the gradients $\nabla_\nu F_{n,0}$ behind the tracer ($n<0$).
 
\subsubsection{Density profiles in the wake of the tracer}
 
We assume that $n<0$ and use the expression of $F_{\boldsymbol{n}}$ from
Eqs.~(\ref{rewriteF}),~(\ref{deff}) and~(\ref{defU}):
 \begin{eqnarray}
\nabla_1 F_{n,0} & = & F_{n+1,0}-F_{n,0} \nonumber \\
 & = & \frac{1}{L} \sum_{k=0}^{L-1} \left[ f(n+1,k)-f(n,k)  \right] \nonumber \\
        & = & \frac{1}{L\At_1} \sum_{k=0}^{L-1}  \frac{U_1(k)-1}{U_1(k)-U_2(k)}\ex{n \ln [U_1(k)]}.
        \label{nabla1_behind}
\end{eqnarray}
The gradient $\nabla_1 F_{n,0} $ has then an exponential
behavior when $n\to-\infty$, dominated by the term of the sum over
$k$ for which $U_1(k)$ is the smallest. 
It can be found by considering the function of a continuous variable $\phi_1$ defined as: $\phi_1(x)=U_1(Lx/2\pi)$, for $x \in [0,2\pi]$. $\phi_1(x)$ is symmetric with respect to the value $x=\pi$, and reaches a maximum for this value. Consequently, 
\begin{eqnarray}
U_1(0)& = & U_1(L) \\
U_1(1) & = &  U_1(L-1) \\
&\ldots& \nonumber
\end{eqnarray}
and the smallest value of $U_1(k)$ is reached for $k=0$ and $k=L$.
We compute $U_1(0)$:
 \begin{equation}
\label{defUbis}
U_1(0) = \frac{1-2\At_2}{2\At_1 }+ \frac{1}{2}\sqrt{\frac{(1-2\At_2)^2}{{\At_1}^2}-4\frac{\At_{-1}}{\At_{1}}},
\end{equation}
and using the relation $\At_1+\At_{-1} + 2\At_2=1$, one obtains
$U_1(0)=1$. The first term of the sum over $k$ in
Eq. (\ref{nabla1_behind}) is then null. We then conclude that the
behavior of $\nabla_1 F_{n,0} $ is dominated by the terms
$k=1$ and $k=L-1$:
\begin{equation}
\label{nab1_behind}
\nabla_1 F_{n,0} \underset{n\to-\infty}{\sim} \frac{2}{L \At_1} \frac{U_1(1)-1}{U_1(1)-U_2(1)} \ex{n \ln[U_1(1)]}.
\end{equation}
With a similar calculation, one obtains
\begin{equation}
\label{nabm1_behind}
\nabla_{-1} F_{n,0} \underset{n\to-\infty}{\sim} \frac{2}{L \At_1} \frac{U_1(1)^{-1}-1}{U_1(1)-U_2(1)} \ex{n \ln[U_1(1)]},
\end{equation}
and
\begin{eqnarray}
\nabla_2 F_{n,0} & = & F_{n,1}-F_{n,0} \\
 & = & \frac{1}{L} \sum_{k=0}^{L-1} \left[ \ex{-\frac{2\ii \pi k}{L}}f(n,k)-f(n,k)  \right] \\
  & = & \frac{1}{L\At_1} \sum_{k=0}^{L-1} \left( \ex{-\frac{2\ii \pi k}{L}}-1 \right) \frac{U_1(k)^{n}}{U_1(k)-U_2(k)}. \nonumber \\
\end{eqnarray}
Once again, the sum over $k$ is dominated by the terms $k=1$ and $k=L-1$:
\begin{eqnarray}
\nabla_2 F_{n,0} &\underset{n\to-\infty}{\sim} &\frac{1}{L \At_1} \frac{\ex{n \ln[U_1(1)]}}{U_1(1)-U_2(1)}\nonumber \\
&\times& \underbrace{ \left(\ex{-\frac{2\ii \pi}{L}}-1+\ex{-\frac{2\ii \pi (L-1)}{L}} -1\right) }_{=2[\cos(2\pi/L)-1]}\\
&\underset{n\to-\infty}{\sim} &\frac{2}{L \At_1} \left(\cos \frac{2\pi}{L} -1 \right)  \frac{\ex{n \ln[U_1(1)]}}{U_1(1)-U_2(1)}. \nonumber \\
\label{nab2_behind}
\end{eqnarray}
Using the asymptotic expansions from
Eqs.~(\ref{nab1_behind}),~(\ref{nabm1_behind}) and~(\ref{nab2_behind})
in Eq.~(\ref{system}), we finally obtain Eqs.~(\ref{h_behind}-\ref{Hm}).
  
\subsubsection{Density profiles in front of the tracer}

We use the expression of $f(n,k)$ for $n>0$:
\begin{equation}
f(n,k) = \frac{1}{\widetilde{A}_1} \frac{{U_2(k)}^{n}}{U_1(k)-U_2(k)}. 
\end{equation}
Studying the properties of the function $\phi_2(x)=U_2(Lx/2\pi)$ on the interval $[0,2\pi]$, one can show that,
depending on the parity
of $L$, the minimum value for $U_2(k)$ is reached:
\begin{enumerate}
  \item  once at $k=L/2$ for $L$ even
  \item twice at $k=\lfloor L/2 \rfloor \pm 1$ for $L$ odd.
\end{enumerate}
For simplicity, we consider the special case where $L$ is even and
obtain
\begin{eqnarray}
  \nabla_{\pm 1} F_{n,0} & \underset{n\to\infty}{\sim} & \frac{2}{L \At_1}  \frac{U_2(L/2)^{\pm1}-1}{U_1(L/2)-U_2(L/2)} \ex{n \ln[U_2(k=L/2)]}\nonumber \\
  \nabla_{2} F_{n,0} & \underset{n\to\infty}{\sim} &\frac{2}{L \At_1} \left(\cos \frac{2\pi}{L} -1 \right)  \frac{\ex{n \ln[U_2(L/2)]}}{U_1(L/2)-U_2(L/2)}, \nonumber 
\end{eqnarray}
and, using Eq. (\ref{system}), we obtain Eqs.~(\ref{h_front}-\ref{Hp}).


\begin{thebibliography}{99}
\bibitem{footnote1} Remarkably, the latter result for a discrete
  system appears to be qualitatively very similar to what was found
  via a dynamic density functional theory~\cite{PDT03} for a driven
  colloidal particle moving in polymer solutions, and has been also
  shown to occur in dense solutions of soft colloids~\cite{DBJ14}.
\bibitem{footnote2} We expect that reflecting boundary conditions do
  not change significantly our results.
\bibitem{footnote2bis} Note that the case $L=1$ corresponds to
  single-files regardless of the dimension $d$.
\bibitem{footnote3} In Ref.~\cite{BM14} the only relevant temporal
  scale is the inverse of bath particles transition rate $\gamma$,
  which is related to our parameters by $\tau^*=1/4\gamma,
  \tau=1/2$. In the simulations reported in~\cite{BM14}, for an
  infinite lattice with density $\rho=0.2$, the phenomenon of negative
  differential mobility is observed for $\gamma\lesssim 0.001$,
  corresponding to $\tau^*\gtrsim 250$.
\bibitem{footnote3bis} Notice that for $\rho\gtrsim 0.5$ and large values
of $\tau^*/\tau$ the times of convergence of numerical simulations are very long 
and more accurate estimations of the TP velocity would require further analysis.
\bibitem{footnote4} Similar behaviors are observed with the choice of
  Eq.~(\ref{maes}).
\end{thebibliography}
\end{document}